\documentclass[preprints,article,accept,pdftex]{mdpi}

\firstpage{1} 
\pubvolume{1}
\issuenum{1}
\articlenumber{0}
\pubyear{2022}
\copyrightyear{2022}
\datereceived{} 
\dateaccepted{} 
\datepublished{} 
\hreflink{https://doi.org/} 
\pdfoutput=1 

\Title{ Dynamical Chiral Symmetry Breaking\\ in Quantum Chromo Dynamics:\\
 Delicate and Intricate}

\TitleCitation{Dynamical Chiral Symmetry Breaking in QCD: Delicate and Intricate}

\newcommand{\dcsb}{D$\chi$SB}

\Author{Reinhard Alkofer}

\AuthorNames{Reinhard Alkofer  \orcidA{0000-0001-7433-3295} }

\AuthorCitation{Alkofer, R.}

\address{
Institute of Physics, University of Graz, NAWI Graz, Universit\"atsplatz 5, 8010 Graz, Austria
}

\corres{Correspondence: reinhard.alkofer@uni-graz.at}

\abstract{
Dynamical Chiral Symmetry Breaking (\dcsb) in Quantum Chromo Dynamics (QCD) for the light quarks is 
an indispensable concept for understanding hadron physics, i.e., the spectrum and the structure of hadrons.
In Functional Approaches to QCD the respective role of the quark propagator has been evident since the 
seminal work of Nambu and Jona-Lasinio has been recast in QCD's terms. It not only highlights one of the most important aspects of \dcsb, the dynamical generation of constituent quark masses, but also makes 
plausible that \dcsb \ is a robustly occurring phenomenon in QCD. The latter impression, however, changes 
when higher $n$-point functions are taken into account. In particular, the quark-gluon vertex, i.e., the most 
elementary  $n$-point function describing the full, non-perturbative quark-gluon interaction, plays a 
dichotomous role: It is subject to \dcsb \ as signalled by its scalar and tensor components but it is also 
a driver of \dcsb \ due to the infrared enhancement of most of its components. Herein, the relevant 
self-consistent mechanism is elucidated. It is pointed out that recently obtained results imply that, 
at least in the covariant gauge, \dcsb \ in QCD is  
located close to the critical point and is thus a delicate effect. And, requiring a precise determination of 
QCD's three-point functions,  \dcsb \  is established, in particular in view of earlier studies, by an intricate interplay 
of the self-consistently determined magnitude and momentum dependence of various tensorial
components of the gluon-gluon and the quark-gluon interactions.}

\begin{document}

\section{Introduction}

Investigations of QCD with the aim of gaining an understanding of hadron physics have been undertaken
since QCD has been formulated almost 50 years ago \cite{Fritzsch:1973pi}. 
The recent review \cite{Gross:2022hyw} summarises on more than 700 pages quite a number of 
highlights arising from these studies. With its almost 5000 references it makes clear how much this 
area of research has matured. Nevertheless, it is agreed upon by the community that the understanding 
of several essential features of QCD and their implications for hadron physics is far from being 
satisfactory. 

In the following short notes 
I focus on a very specific property of QCD, namely the approximate chiral symmetry
of the light quarks and how it is dynamically broken. 
Despite the importance of \dcsb \ for the phenomenological 
consequences with respect to the spectrum and structure of hadrons I am concentrating herein on the
underlying mechanisms for \dcsb, or more precisely, on a detailed analysis within the picture that a 
super-critically strong attraction in between massless fermions triggers \dcsb, see, e.g., 
\cite{Miransky:1979ks,Fomin:1983kyk,Miransky:1994vk}. 
To this end one may note that more than 60 years ago 
Nambu and Jona-Lasinio realised that in four spacetime dimensions a certain coupling strength has 
to be exceeded for \dcsb \ to occur \cite{Nambu:1961tp,Nambu:1961fr}. 

At this point a disclaimer is in order: Herein, I will summarise and briefly review some investigations 
of \dcsb, the choice of which is based on my own attempts within this field of research. By no means 
it is intended to disregard different approaches to the topic which are based on complementary
techniques and/or pictures (as, e.g., an explanation of Chiral Symmetry Breaking by considering 
ensembles of QCD vacua containing lumps of gluon fields with non-vanishing
topological winding number densities). And given the wealth of literature on \dcsb, even if one 
restricts oneself (i) to the picture of a super-critically strong attraction as underlying mechanism 
and (ii) to functional methods, it is impossible within such a short synopsis as the one presented here 
to discuss or even mention all relevant research on this topic. Such omissions are also in accord with the 
intention of the presented discussion: To provide evidence that \dcsb \ in QCD is quite delicate, its 
manifestations in the properties of quarks and quark-gluon interactions exhibit many facets, and the
interplay in between those features makes \dcsb \  an intricate idiosyncrasy of QCD.

\section{How robust is \dcsb \ in QCD?}

\subsection{The Nambu--Jona-Lasinio picture}

The seminal papers by Nambu and Jona-Lasinio  \cite{Nambu:1961tp,Nambu:1961fr} introduced
the notion of \dcsb \ in analogy to the shortly before formulated BCS model of superconductivity 
\cite{Bardeen:1957kj,Bardeen:1957mv}. The generic idea of Nambu and Jona-Lasinio  has been 
that massless (light) nucleons interact via a four-fermion interaction which in turn leads to massive 
nucleons and (almost) massless pions as (pseudo-) Goldstone bosons. As their starting point was
to describe nucleons as massless Dirac fermions interacting via a SU$_L$(2) $\times$  SU$_R$(2) 
chirally invariant interaction the dynamics of their model respects chiral symmetry but the ground 
state symmetry was broken down to a vector SU$_{L+R}$(2) symmetry. Therefore, given the
three-dimensional coset space, three pseudo-scalar massless, resp., light excitations arise as 
(would-be) Goldstone bosons, the pions. 

From this one can take away three important lessons:
\begin{itemize}
\item In contradistinction to spontaneous symmetry breaking the mechanism of dynamical symmetry
breaking introduces a dichotomous nature for the (would-be) Goldstone bosons, they are not only
Goldstone bosons but at the same time bound states of a highly collective nature. This is true for 
the original picture based on nucleons but, of course, also if one starts with light quarks interacting 
at the tree-level in a chirally symmetric way, see, e.g., the discussion in \cite{Alkofer:1995mv}.
\item \dcsb \ implies the generation of dynamical masses for originally massless and/or light fermions. 
This solves the puzzle why in the quark model for the light quarks the so-called constituent quark 
masses at the order of 
$\gtrsim$ 350 MeV are required instead of the much smaller current quark masses. 
\item In contradistinction to non-relativistic superconductivity where Cooper pairs are formed
at arbitrary small couplings\footnote{As a matter of fact, this statement is only true in the 
mean-field approximation. When taking into account fluctuations also a certain minimal coupling
is required to form Cooper pairs.} \cite{Bardeen:1957mv} \dcsb\ in four spacetime dimensions only takes
place if the coupling exceeds a critical value. 
\end{itemize}

\begin{figure}
\begin{center}
\includegraphics[width=0.6\textwidth]{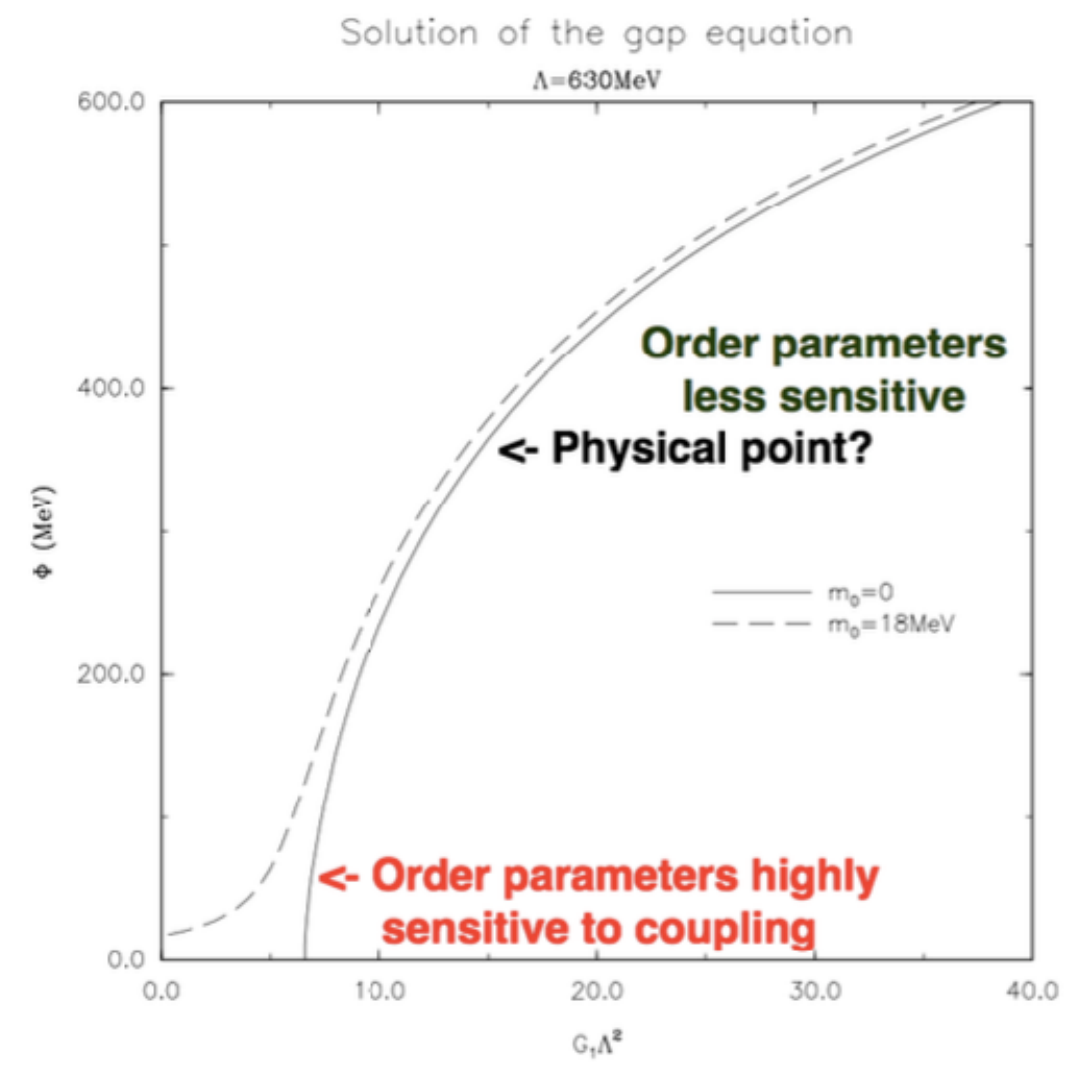}
\end{center}
\caption{\label{FigMvsG} An example for the generated constituent quark mass as a function 
of the coupling within a Nambu--Jona-Lasinio model calculation. (Adapted from 
Ref.~\cite{Alkofer:1995mv}.)}
\end{figure}

Although these three statements are correct, they alone provide an incomplete picture. 
Before explicating in which sense the second statement has to be augmented it is instructive to have a closer look at the third one. In the chiral limit any order parameter will show as a function
of the coupling a non-analyticity at the critical value of the coupling. For light quarks, i.e., in the case of 
approximate chiral symmetry, one has a cross-over characterised by a rapid change of the would-be 
order parameter. This is illustrated in Fig.~\ref{FigMvsG} in which for a calculation within a 
Nambu--Jona-Lasinio model the constituent quark mass is shown as a function of the four-fermion 
coupling, for the details see Ref.~\cite{Alkofer:1995mv}. This calculation seems to imply that the 
physical point is such that the corresponding coupling is much larger than the physical one, and 
correspondingly all order parameters would depend only mildly on the precise value of the coupling. 
In the following I will argue that this behaviour seen in a Nambu--Jona-Lasinio model (and certain 
truncations to QCD)  is not correct for QCD. The most important effect of this is the resulting sensitivity 
of all chiral order parameters on the precise value of the quark-quark interaction strength.

\subsection{On the Dyson-Schwinger / Bethe-Salpeter approach in Rainbow-Ladder truncation}

As \dcsb \ is a non-perturbative phenomenon methods beyond perturbation theory are needed 
to investigate it. If it comes to the study of dynamical symmetry breaking an approach based 
on Dyson-Schwinger and Bethe-Salpeter equations has been widely employed, see, e.g., the
textbook \cite{Miransky:1994vk} for an introduction. In particular, this approach has been used 
widely in the context of QCD and hadron physics as documented by a number of reviews 
\cite{Roberts:1994dr,Roberts:2000aa,Alkofer:2000wg,Maris:2003vk,Bashir:2012fs,Eichmann:2016yit,Sanchis-Alepuz:2017jjd}.

An essential element in this approach is the choice of a symmetry-preserving truncation of 
the infinite set of equations for the $n$-point correlation functions. Within a Poincar\'e-covariant
setting (implying, at least implicitly, the choice of a covariant gauge, cf. the discussion below
in sect.~\ref{SecGauge}) the simplest non-trivial 
of such approximations is the rainbow-ladder truncation. It owes its name because the 
infinitely many re-summed  diagrams look like rainbows for the quark propagator's 
Dyson-Schwinger equation and like ladders for the mesons' bound state equations, the 
Bethe-Salpeter equations. 

Since Ref.~\cite{Fukuda:1976zb} several hundred solutions of the quark propagator's 
Dyson-Schwinger equation in rainbow approximation have been published, 
and since Ref.~\cite{Jain:1991pk,Munczek:1991jb} a similar number  of solutions for the pion
 Bethe-Salpeter equation in ladder approximation have been described in the literature. 
For many but not all hadrons such an approximation works astonishingly well, 
see, e.g., Ref.~\cite{Eichmann:2016yit} for a detailed discussion and 
Refs.~\cite{Williams:2015cvx,Vujinovic:2018nko,Miramontes:2021xgn,Gao:2021wun,Miramontes:2022mex} 
for some examples of beyond-rainbow-ladder calculations.\footnote{For a study of the 
functional renormalisation group taking a dynamical quark-gluon vertex into account, see, e.g.,
 \cite{Cyrol:2017ewj}.}

For the purpose of these notes the use of the rainbow-ladder truncation will not allow
to resolve the issues raised in the preceding section. The reason is quite simple: In this
truncation the quark-gluon vertex is given by a model, and for technical reasons the 
employed models are incomplete. Important aspects of the effect of \dcsb \ on the 
quark-gluon interaction are thereby excluded by assumption. Phrased otherwise, one cannot find
what one excludes by approximation.

A further ``twist`` of the rainbow-ladder truncation lies in the reduction to the tree-level tensor component
in the quark-gluon vertex followed by a fitting of the overall interaction strength to phenomenological data. 
This then leads, as argued in the next section, to an overestimate of the coupling strength between
quarks and gluons.

\subsection{On the onset of the Conformal Window}

It is evident from hadron phenomenology that \dcsb \ takes place in QCD. For a Gedankenexperiment
let us consider a gauge theory with $N_f$ massless (or light) fermions in the fundamental representation
of the gauge group. If $N_f$ is small the anti-screening caused by the gauge bosons dominates, and
consequently the running coupling will increase when tuning the scale from larger to smaller scales.
Eventually, it will exceed the critical coupling, and \dcsb \ will take place. At very large $N_f$ the screening
caused by the fermions will dominate, and asymptotic freedom will be lost. 

However, in between this two extremes there will exist an interval for $N_f$ where the balance in between
the anti-screening due to the gauge bosons and the screening due to the fermions is such that the 
anti-screening effect wins only so slightly against the screening. Correspondingly, the coupling will increase
when lowering the scale but only so weakly that the critical coupling is never exceeded. Thus, \dcsb \ 
will not take place. In the chiral limit, such a theory possesses an infrared fixed point, it will be effectively 
scale-invariant in the deep infrared. For that reason the corresponding interval for $N_f$ is called the 
conformal window.

Although the above described generic picture has been verified by studies based on coupled 
Dyson-Schwinger equations \cite{Hopfer:2014zna,Fabian}, however, the critical value for the numbers of flavours at 
which the conformal window sets in, $N_f^{crit}$, is severely underestimated when compared to studies employing 
other methods, see, e.g., \cite{DeGrand:2015zxa,Witzel:2019jbe,Gies:2005as,Lee:2020ihn}.
The decisive hint why Dyson-Schwinger studies in rainbow-ladder truncation show such a deficiency 
comes from the sensitivity of $N_f^{crit}$ on the quark-gluon vertex if one goes (slightly) beyond the 
rainbow-ladder truncation.  This behaviour makes plain that the distribution of the overall quark-gluon interaction strength
in the sub-GeV region over several of the quark-gluon vertex tensor structures, as it happens without any doubt in QCD, 
is essential in an understanding how the increased screening by an increasing number of massless, resp., light, quark 
flavours drives the system into a chirally symmetric phase with an IR fixed point.

\subsection{A note on gauge dependence}
\label{SecGauge}

Since the seminal work by Curtis and Pennington \cite{Curtis:1990zr} it has become evident how important the 
fermion--gauge-boson vertex is in achieving gauge independence in the Dyson-Schwinger approach. Although for QED 
substantial progress has been achieved, see, e.g., 
\cite{Jia:2016aau,Albino:2022efn,Guzman:2023hzq} and references therein, in the 
studies of the role of the fermion-photon vertex for gauge independence the corresponding question in QCD , namely on the
impact of the quark-gluon vertex on the gauge (in-)dependence of hadron observables, has proven to be an extremely 
hard question. Even the much more humble question how the different tensors of the quark-gluon vertex may depend
on the gauge parameter within the class of linear covariant gauges and how this will effect the underlying mechanism for \dcsb \ in this class of gauges seems beyond reach given the status of  Dyson-Schwinger studies of the Yang-Mills sector
in the linear covariant gauge, see, e.g., \cite{Aguilar:2015nqa,Aguilar:2016ock,Napetschnig:2021ria}. 

Therefore, although the question whether \dcsb \ in QCD is delicate and intricate only in the Landau gauge and might be a 
robust phenomenon in other gauges is highly interesting it will likely remain to be unanswered in the next years. 
Nevertheless, 
in view of the insights which may be gained in studying the role of the quark-gluon vertex and its impact on \dcsb \
in different gauges an extension 
of the approach based on Nielsen identities (as performed in \cite{Napetschnig:2021ria}) to the quark sector is certainly 
desirable. One might also apply the technique of interpolating gauges 
\cite{Fischer:2005qe,Alkofer:2003jr,Capri:2006bj,Andrasi:2021qrw}
to relate the existing Landau and Coulomb gauge results on \dcsb . Until such studies will succeed the herein described
analysis will only be applicable to QCD in the Landau gauge. 

\bigskip \bigskip 
  
\section{Correlation functions in the Yang-Mills sector}

In order to investigate the interplay between the quark propagator and the quark-gluon vertex within functional methods one
needs to be able to determine the propagators and the three-point functions in the Yang-Mills sector accurately. In the last two
decades there have been enormous progress in this direction, see, e.g., the reviews
\cite{Huber:2018ned,Huber:2020keu,Ferreira:2023fva}, and it is fair to say that in the Landau gauge 
the gluon and the ghost propagators as well as the three-gluon and the ghost-gluon vertex are well understood.

Hereby, two features are important. 

First, the gluon propagator's renormalisation function as function of the gluon virtuality 
$p^2$ displays on the space-like side a maximum slightly below one GeV, and then decreases towards the infrared. This 
unusual behaviour not only signals a strongly reduced spectral dimension  \cite{Kern:2019nzx} and relates the 
gluon long-range properties to non-vanishing $p^2$  \cite{Fischer:2005ui,Kern:2019nzx} but it also leads to the fact that 
the gluon propagator alone, i.e., without quark-gluon vertex dressings, is much too small in the sub-GeV region to trigger
\dcsb , see, e.g., the discussion in \cite{Fischer:2003rp}.

Second, the three-gluon vertex gets suppressed towards the infrared, and the corresponding form factors display 
in the most accurate available calculations even a zero at small values of $p^2$. As in the Dyson-Schwinger equation for the 
quark-gluon vertex the three-gluon vertex turns out to be decisive in determining the infrared enhancement of the 
quark-gluon vertex form factors which in turn determines the size and the proximity to criticality of \dcsb \ these two 
observations together explain why in QCD in Landau gauge \dcsb \ is so delicate in distinction from models ignoring these 
two facts. 

\section{Quark propagator and quark-gluon vertex}

\subsection{Structure of the quark-gluon vertex}

The arguments provided above elucidate the special role of the quark-gluon vertex in the description of \dcsb \ in the
Landau gauge. Unfortunately, this vertex possesses a rich structure, and it is exactly the interplay in between parts of this
structure which turn out to be relevant for the physics of \dcsb .

There is one straightforward property of the fully dressed quark-gluon vertex: To the best of our knowledge it possesses 
the same colour structure as its tree-level counter part.

When it comes to flavour, and in particular to the dependence on the current quark mass, a careful assessment of the
properties of the substructures is in order. To this end one notes first that in the Landau gauge only that parts of the 
vertex are relevant which are strictly transverse to the gluon momentum. As the quark-gluon vertex transforms as four-vector 
under Lorentz and as a Dirac matrix under spin rotations this leaves one with eight possible tensor 
structures, each tensor structure being multiplied with a form factor depending on three Lorentz-invariant variables 
which in turn are built from the three involved momenta.

Instead of choosing immediately a definite basis for this eight tensors it is worthwhile to discuss some generic 
aspects first. The Feynman integrals for the form factor multiplying the tree-level tensor are ultraviolet divergent, and 
thus this one form factor needs renormalisation. Choosing the other seven tensors orthogonal to the tree-level one
the corresponding form factors are determined from ultraviolet-finite expressions, and correspondingly they decrease 
power-like for large momenta. This leads to the expectation, later on confirmed by calculations, 
that these form factors are only sizeable if at least one of the involved momenta is small, i.e., in the sub-GeV region. 

The eight tensors of the transverse part of the quark-gluon vertex can grouped according to their behaviour under 
chiral transformations: Four of them are chirally symmetric, and thus they will be generically non-vanishing even in 
the chiral limit and the symmetric Wigner-Weyl phase of chiral symmetry. In that latter case the form factors of the other 
four chirally non-symmetric tensor structures vanish. In the Nambu-Goldstone phase they will be dynamically generated, and
phrased otherwise this exactly means that \dcsb \ also includes the generation of chirality-violating scalar and tensor 
quark-gluon interaction. As can be seen below they are quite sizeable.

If it comes to the dependence of the quark-gluon vertex on the current quark mass, i.e., on the explicit breaking of chiral 
symmetry, this distinction in between the chirally symmetric and non-symmetric parts lead to a quite astonishing behaviour 
of the latter components. The Feynman diagrams for the quark-gluon vertex contain at least one quark propagator within
a loop. Of course, this quark propagator goes to zero as the quark mass goes to infinity. Therefore, naively one 
might conclude that the fully dressed quark-gluon vertex will approach the tree-level one for larger and larger current 
quark masses. However, one has to take into account that the chirally non-symmetric form factors by the mere virtue of 
their transformation properties also will have a factor of at least one current quark mass and/or dynamically generated 
constituent quark mass in the numerator. Therefore the suppression by the current quark mass in the denominator of the 
integrand introduced via the quark propagator can and generically will be canceled.

As a matter of fact, this mechanism is already at work in QED w.r.t. the Pauli term and the resulting anomalous magnetic
moments (g-2): There is a cancelation of factors of the fermion mass in the QED contributions to, e.g., the (g-2) of the 
electron and the muon. 

\subsection{Dynamical generation of scalar and tensorial quark-gluon interactions} 

In the following the above statements will be quantified on the basis of the results obtained in \cite{Windisch:2014lce}, 
see also \cite{Hopfer:2012cnq,Alkofer:2013qoc,Blum:2016fib}.\footnote{The 
interested reader will find figures of the quark-gluon vertex' form factors in these references.}
 I want to emphasise here that the corresponding results 
of other groups would have been equally valid, the choice is only based on the availability of the data. And to be concise
within this short note only results in the chiral limit will be discussed. 

The following kinematics is chosen:

Gluon momentum: $k^\mu = p^\mu - q^\mu$ with $p^\mu$ outgoing and $q^\mu$ incoming quark momentum.

Define furthermore:

(i) Normalised gluon momentum: $$\hat k^\mu := k^\mu / \sqrt{k^2}.$$

(ii) Averaged quark momentum, $\frac 1 2 (p^\mu + q^\mu)$, project it transverse to gluon
momentum and normalise it
$$
s^\mu := (\delta^{\mu\nu} - \hat k^\mu \hat k^\nu ) \frac 1 2 (p^\nu + q^\nu) \, ,
\quad \hat s^\mu = s^\mu / \sqrt{s^2}\, .
$$

As a three-point function the quark-gluon vertex, or more precisely the factors
multiplying the tensors in a decomposition, depend on three Lorentz
invariants, and we choose them to be $p^2$, $q^2$ and $p\cdot q$.
The transverse part of the quark-gluon vertex is expanded then in the form
\begin{equation}
\Gamma^\mu_{trans} (p,q;k) = \sum_{j=1}^8 g_i (p^2,q^2;p\cdot q) \rho_i^\mu
\, , \label{QGV}
\end{equation}
and in the following we will approximate the transverse part of the
quark-gluon vertex. First, as the
angular dependence turns out to be weak we will neglect it.
The functions $g_i (p^2,q^2;p\cdot q)$ are symmetric in $p$ and $q$,
therefore we will substitute them by functions $g_i(\bar p^2)$ of only the
averaged momentum-squared, i.e., $\bar p^2 =
\frac 1 2 (p^2+q^2)$.
The model functions  $g_i(\bar p^2)$ are fitted to the numerical results at
symmetric momenta, $g(p^2,p^2; p\cdot q =0)$ obtained from a coupled set of
quark propagator and quark-gluon vertex
Dyson-Schwinger equations in the chiral limit with a model
for the three-gluon vertex, see  
\cite{Windisch:2014lce,Hopfer:2012cnq,Alkofer:2013qoc,Blum:2016fib} for more details.

Hereby it turns out that $g_1$, $g_2$, $g_3\propto g_2$ and $g_4=g_7$ are
important whereas based on the underlying results for $g_5$ and $g_8$
it is safe to neglect these two functions.

Employing that to numerical accuracy $g_4=g_7$, and that one observes
$g_3\propto g_2$ in the sense that $ 1.45 \, g_2(p^2,p^2,0) + g_3(p^2,p^2;0)$ is
for all momenta smaller than 0.08, one is left {\bf with effectively three tensor structures}.

1. Tree-level tensor structure (with $x=\bar p^2 /$ 1 GeV$^2$):

$\rho_1^\mu = \gamma^\mu_T = (\delta^{\mu\nu} - \hat k^\mu \hat k^\nu )
\gamma^\mu$, with 

$g_1(\bar p^2) = 1 + ( 1.6673 + 0.2042 x)/(1+0.6831 x + 0.0008509 x^2)$

Of course, the tree-level tensor structure is allowed in the chirally symmetric
phase.

2. The further sizeable chirally symmetric tensor structure is given by:

$\rho_4^\mu +
\rho_7^\mu  = \hat k \!\!\!\! / \,\, \hat s^\mu  +\hat s \!\!\! / \,\, \hat k \!\!\!\! / \,\, \gamma^\mu_T$, with 

$g_4 (\bar p^2) =g_7(\bar p^2) = 2.589 x / ( 0.8587 + 3.267 x + x^2)$

3. The one important tensor structure due to (dynamical or explicit) chiral symmetry breaking is
a combination of
$\rho_2^\mu = i \hat s^\mu $ and $\rho_3^\mu = i  \hat k \!\!\!\! / \gamma^\mu_T$.

The corresponding form factors are
$g_3(\bar p^2) = 0.3645 x / ( 0.01867 + 0.3530 x + x^2)$,
$g_2(\bar p^2) = - g_3(\bar p^2) / 1.45$, and the latter relation also fixes the relative weight 
of the 2nd and the 3rd component in the expansion (\ref{QGV}).

Hereby, $\rho_2^\mu$ is a Dirac scalar (i.e., proportional to the unit matrix), and
$\rho_3^\mu$ a rank-2 tensor.

Therefore, the one main conclusion of this section is that in QCD in Landau gauge 
a {\bf scalar and a tensorial quark-gluon interaction} is dynamically generated. 
Phrased otherwise, non-perturbatively fully dressed gluons interact with quarks as
if they had a spin-0 and spin-2 component. 

\subsection{The coupled system and its lessons for \dcsb}

Putting all the above pieces together one realises that a description of \dcsb \ in QCD 
in Landau gauge and based on the fully dressed quark, gluon and ghost propagators 
as well the fully dressed three-point functions displays quite an elaborate web of 
self-consistent interdependencies. Contrary to what has been assumed in the early days
of QCD, namely that the gluon propagator is the main driver of a robust version of \dcsb , it turns out that 
the intricate interplay between all the involved functions puts the whole system close to 
criticality. Although amongst these functions the quark-gluon vertex is the richest in structure
it is the one quantity which allows to improve on our understanding of the complicated way
the fully dressed gluon interacts with fully dressed quarks in the strongly interacting domain.

From a bird's eye perspective this should not come as a surprise. It is obvious from the 
experimental results in hadron physics that thresholds which are apparent in 
scattering cross sections stem from intermediate hadron resonances. Despite its rich
structure the quark-gluon vertex is still the simplest among all the QCD correlation functions
which could seed such dependencies. Together with an understanding how the kinetic terms 
for hadrons might emerge from the QCD degrees of freedom (for a corresponding discussion, see, 
e.g., \cite{Paris-Lopez:2018nua}) this opens up the possibility to map out the wealth of hadron 
physics with less than a dozen functions derived from QCD. Therefore, the richness 
of these functions and of the equations determining them should not come as
a surprise.

\section{Conclusions and Outlook}

In this short note I argued that the view on \dcsb \ in QCD needs to take into account the results
obtained over the last two decades for the correlation functions of gluons and quarks. Having 
been seduced by some older results to believe that \dcsb \ in Strong Interactions is a robust 
phenomenon (due to the reason that interactions are strong) the more recent results urge us
to re-think this point of view: It looks much more that \dcsb \ is delicate and intricate.

At this point one might argue that this distinction between robust \& straightforward vs. \
delicate \& intricate might only be an interpretational one. To my opinion there are at least three 
reasons to pay attention to the view advocated here. The first one is within hadron physics
itself. Being aware about the sensitivity in the description of  \dcsb \ provides some guidance
in understanding which hadron observables will inherit this sensitivity on the details of the 
underlying quark and glue dynamics. In this respect the question of the formation of a hadron
provides quite likely one of the main examples of an intricate process. Second, quite a number
of models beyond the Standard Model
as, e.g., technicolor, exploit a potential proximity to the lower end of the conformal 
window to generate a ``walking'' coupling and correspondingly a vast separation of scales. Needless
to say that an understanding of the transition to the conformal window and the physics therein (as 
well as close to it) will build on the details of the fate of chiral symmetry in this parameter domain.
Third (but not least), I'd like to remind the reader that the Standard Model possesses another 
chiral transition triggered by the Higgs-Yukawa couplings and happening at the electroweak scale.
(Some insight into how intricate  these two chiral transitions intertwine 
can be inferred from the recent investigation reported in ref.\ \cite{Gies:2023jzd}).
Therefore, a deepened insight into the chiral properties of the Standard Model fermions will always
need to include the very nature of \dcsb \ within QCD.

\bigskip \bigskip 

\acknowledgments{It is a pleasure to cordially thank all colleagues who collaborated with me 
on the topics presented here. I am in particular grateful for the
insights gained in my many  respective  discussions with
Per Amund Amundsen, William Detmold, Gernot Eichmann,  Christian Fischer, Markus Hopfer, 
Markus Huber, Felipe Llanes-Estrada, Axel Maas, Pieter Maris, Angel Miramontes, Mario Mitter, 
Jan Pawlowski, Hugo Reinhardt, Alexandre Salas-Bernardez, 
Helios Sanchis-Alepuz, Lorenz von Smekal, Milan Vujinovic, 
Herbert Weigel, Richard Williams, Andreas Windisch, Fabian Zierler and Daniel Zwanziger.}

\conflictsofinterest{The author declares no conflict of interest.}

\bigskip \bigskip

\bigskip

\begin{adjustwidth}{-\extralength}{0cm}

\reftitle{References}

\end{adjustwidth}
\end{document}